# Investigation of Tactile Texture Simulation on Online Shopping Experience

Pei Hsin Lim[1], and Kian Meng Yap[1]

[1] *Research Centre for Human-Machine Collaboration (HUMAC), School of Engineering and Technology, Sunway University, No. 5, Jalan Universiti, Bandar Sunway, 47500 Selangor Darul Ehsan, Malaysia.*

(Email: limpeihsin15@gmail.com, kmyap@sunway.edu.my)

**Abstract ---** With safety measures towards the current Covid-19 pandemic, many retails clothing stores have restricted on-site fittings and shifted their business online. Inability to touch on product evaluations shows an apparent limitation as compared to retail shopping especially when the object's material information is crucial like clothing. Haptic technologies show potential of bridging the gap between online shops and the shoppers by providing a sense of touch, yet little research has been done especially on the effect of the simulation of tactile texture on the shopping experience. In this study, we modified a mock-up e-commerce website by adding clothing products and enabling a mid-air haptic interface with Ultrahaptics Evaluation Kit (UHEV1). We developed texture sensations using Time Point Streaming (TSP) modulation for clothing products with different texture materials and a user study was carried out to investigate the tactile texture sensation on shoppers' experience in evaluating online products. Our results show that tactile texture sensation using multipoint mid-air haptic feedback improves online shopper's satisfaction on the product browsing experience. This study contributes to the improvement of general lifestyle of the society in terms of e-commerce experience and could expand its application to impact different sectors like education and different communities including the visually impaired.

Keywords: haptics, haptic feedback, user experience, user interface, e-commerce

## 1 INTRODUCTION

With safety measures towards the current Covid-19 pandemic, many retails clothing stores have restricted on-site fittings and shifted their business online. With increasing growth of online shopping, the consequences of inability to touch on product evaluations shows an apparent limitation as compared to retail shopping especially when the object's material information is crucial like clothing.

Haptic technologies show potential of bridging the gap between online shops and the shoppers by providing a sense of touch [1], yet little research has been done especially on the effect of the simulation of tactile texture on the shopping experience. A recent mid-air ultrasound haptic display technology uses multiple points of haptic interface points enabling reproduction of macro and micro level texture simulation by generating ultrasonic waves in mid-air [2, 3, 4, 5, 6, 7, 8, 9, 10]. With that, the question in point will be how realistic the simulation touch will be, and does it compensate for the lack of touch in the online shopping experience?

In this paper, we modified a mock-up website created in our previous work by adding clothing products and enabling haptic interface with a mid-air haptic device, Ultrahaptics Evaluation Kit (UHEV1) to study whether the haptic device with tactile texture can help to enhance shoppers' experience in evaluating online products. To enable consistent browsing experience, we align the haptic model with 3D graphic models on the website to provide both touch and visual cues of the products.

## 2 RELATED WORK

Zhang et al. [2] conducted a user study using single point of interface haptic devices, Geormagic® Touch and Novint® Falcon, to compensate for the missing in haptic feedback in enhancing online shopping experience.

Beattie et al. [4] presented a method to simulate haptic dimension of texture in virtual and holographic objects using mid-air ultrasonic technology (Ultrahaptics STRATOS Inspire Device). Freeman et al. [5] rendered roughness in mid-air ultrasound haptics. However, both [4] and [5] did not involve human subject experiments to demonstrate their prototype and the effect of the prototype on users. In a more recent study by Beattie et al. [6], they integrate visual perception of roughness into the rendering algorithm of mid-air haptic texture feedback.

Takahashi et al. [7] proposed a new tactile ultrasound stimulation method with lateral modulation (LM) as opposed to the conventional amplitude modulation (AM). Ablart et al. [8] introduced a new modulation technique called Spatiotemporal Modulation (STM) to simulate lines, curves, and shapes by moving a mid-air tactile point rapidly and repeatedly along the path.

Sand et al. [10] combined mid-air tactile feedback with fogscreen display to present a mixed reality prototype in thin air. They conducted a user test with 12 participants to investigate the tapping performance with and without tactile feedback. Their results show no statistically significant difference between the two groups.

## 3 METHODOLOGY

### 3.1 System Development

This project inherits a haptics enabled and W3C Web Accessibility compliant e-commerce website named Jomje (link: http://jomje.homeislab.com/) from [13]. A new Fashion category with 3 products named Jeans, Hoodie, and long sleeves T-shirt are added into the website. 3D model X3D files are downloaded from online sources [11,12] and implemented with H3DAPI for stylus-based haptic interaction. A new "Feel me" button is added in the Product details page to download mid-air haptic executable file.

The haptic sensation for the UHEV1 model is developed using Ultrahaptics SDK (version 2.6.2) and Leap Motion SDK (version 4.1.0) with C++ and cmake according to guidance from Ultrahaptics official website [14]. UHEV1 is made up of a 16x16 array of ultrasonic speakers that simulates sensation via ultrasonic waves. UHEV1 supports two modes of operations: Amplitude Modulation (AM) and Time Point Streaming (TPS).

In this study, TPS modulation was selected over AM due to its flexibility in controlling the modulation of the control point (CP) amplitude and position [15]. This allows spatiotemporal modulation that was used in [8] and [9] that creates smoother and quieter perceptible waves with lower latency [16]. Three types of texture sensation were developed to represent the clothing products chosen. The control points were set to be 20 cm above the array for best performance and a square mesh was modulated to represent the surface of the clothing. A square mesh was chosen for simplification as this study focuses on the tactile texture experience.

The three types of modulation were developed: TPS_ST, TPS_STLI, and TPS_S. Each type modulates a square mesh that gives different textures by varying their y-offset and z-offset for their CP positions, intensity, and frequency as shown in Table 1. A -0.1 cm z-offset was set for alternating CP to simulate the roughness of the texture.

Table 1 The variables set for each Texture modulations

| variable | side length [cm] | intensity | frequency [hz] |
|---|---|---|---|
| TPS_ST | 3 | 1 | 100 |
| TPS_STLI | 3 | 0.5 | 100 |
| TPS_S | 3 | 1 | 120 |

### 3.2 Experiment Setup

Figure 2 shows the experiment setup consisting of a laptop PC with access to the Internet connected to a mouse and wired to two haptic devices: mid-air haptic device which consists of UHEV1 and a Leap Motion controller with a 20cm height hand rest, and stylus-based haptic device, Geomagic TOUCH. The task list and Jomje Website was opened using the laptop. The hand rest was made to control the height of the user's hand precisely at 20 cm above the array for consistent experience.

The overall connection schematic is shown in Figure 3. The laptop accessed the Internet through Wi-Fi connection to access the Jomje server via wireless access. UHEV1 and leap motion controller were connected to the laptop via a USB cable. Geomagic TOUCH were connected via Ethernet cable.

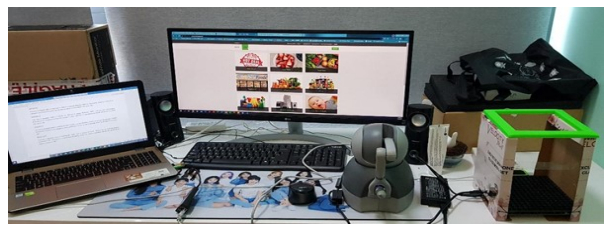

Fig.1 Experiment Setup

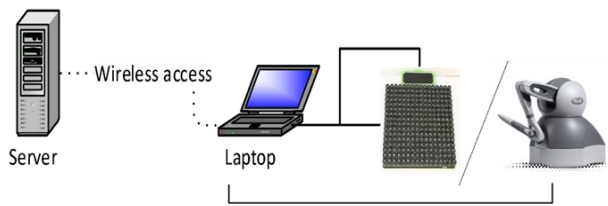

Fig.2 Schematic connections of the haptic devices, motion controller, laptop, and server

## 3.3 User Study

Participants were selected to experiment online shopping with conventional interface (mouse) and mid-air haptic interface (tactile feedback). The former being the control group and the latter being the experiment group. We have decided to use correlated t-tests, with control and experimental group being the same sample, before and after. After the experiment, the participants will be required to fill in a questionnaire on the user experience. The results of the questionnaire will undergo one-tail correlated t-test analysis to conclude if the hypothesis is accepted.

Null hypothesis 1: There is no difference between the users' satisfaction on online shopping experience for both control group and experimental group.

Hypothesis 1: Multipoint mid-air haptic feedback improves online shopper's satisfaction on the product browsing experience.

Apart from comparing between conventional interface and mid-air haptic interface, comparison of user experience between stylus-based single point haptic interface and mid-air haptic interface is also carried out as an extension from previous study [13]. The former being the control group and the latter being the experiment group. The following hypotheses were tested:

Null hypothesis 2: There is no difference between the users' satisfaction on online shopping experience for both control group and experimental group.

Hypothesis 2: Multipoint mid-air haptic feedback improves online shopper's satisfaction on the product browsing experience better than single-point stylus-based haptic feedback.

A seven-point Likert scale [17] is used to capture the degree of agreement of the user for each statement. Likert scale is chosen because quantitative measurement on the experience of the user is needed for the analysis, and it is one of the widely used rating scales to measure attitudes [18].

Questionnaires:

1. Compared to in-store shopping, my understanding of the texture of the given products through this interface can be rated as: 1-very poorly; 7-as good as in-store shopping.

2. Compared to in-store shopping, I find this browsing experience satisfying: 1-strongly disagree; 7-strongly agree.

3. I can compare and distinguish different types of texture through this interface: 1-strongly disagree; 7-strongly agree.

4. Understanding of the texture of the given products through this interface affects my purchase decision: 1-strongly disagree; 7-strongly agree.

## 4 RESULTS AND DISCUSSION

Three users who are daily online shoppers participated in this study, with each of them carried out the tasks in four scenario settings in numerical order. Scenario 1 represents the control group (mouse interface), and Scenario 2 represents the experiment group (mid-air haptic interface) for Hypothesis 1. Scenario 3 represents the control group (stylus-based haptic interface), and Scenario 4 represents the experiment group (mid-air haptic interface) for Hypothesis 2. The correlated t-test was computed to analyse the statistical difference between S1 and S2, and S3 and S4 and the results from the questionnaires are shown in Table 2.

The result shows that at $p<0.025$, the rating difference between S1 and S2 was significant. Hence, we reject Null Hypothesis 1 and accept Hypothesis 1 at confidence level 97.5%. The result shows that at $p<0.05$, the rating difference between S3 and S4 was insignificant. Hence, we reject Hypothesis 2 and accept Null Hypothesis 2. Although there was an increase in the mean values from S3 to S4, there was not enough evidence to prove that the difference was significant.

Through oral feedback, we found that 2 out of 3 participants could differentiate the texture sensation of different clothing products with a mid-air haptic interface. All participants mentioned that the texture sensation in stylus-based haptic interfaces was not obvious; they could tell the shape but not the texture of the surface of the products. With a mid-air haptic interface, they could sense the texture better and it is more natural to feel with bare hands.

Table 2  Average ratings of the overall satisfaction of the users and the t-test results

|  | Test for Hypothesis 1 | Test for Hypothesis 2 |
|---|---|---|
| Control Mean | 4.4167 | 5.1667 |
| Experiment Mean | 5.6667 | 6.0833 |
| t-value | 2.529 | 1.733 |
| p-value | 0.025 | 0.05 |
| t-critical value | 2.201 | 1.796 |
| Significance | Significant | Insignificant |

## 5 CONCLUSION

In this study, we modified a mock-up website created in [13] by adding clothing products and enabling a mid-air haptic interface with UHEV1. We developed 3 different texture sensations using Time Point Streaming Modulation for 3 clothing products with different texture material. A user study was carried out to investigate if a haptic interface with tactile texture enhances shoppers' experience in evaluating online products. Based on the results, we conclude that multipoint mid-air haptic feedback improves online shopper's satisfaction on the product browsing experience. In addition, we also developed 3 shapes from the existing products to investigate the difference in experience of the existing stylus-based haptic device (Geomagic TOUCH) and our newly implemented mid-air haptic device (UHEV1). Our study found that there is an increase in the mean for the overall experience ratings from stylus-based haptic interface to mid-air haptic interface, however, there was not enough evidence to prove that the difference was significant.